\documentclass[12pt,amssymb, amsfonts]{article}
\usepackage{graphicx, amssymb}
\usepackage{algorithmic}
\usepackage{algorithm}
\usepackage{authblk}

\newtheorem{theorem}{Theorem}[section]
\newtheorem{corollary}[theorem]{Corollary}
\newtheorem{definition}[theorem]{Definition}
\newtheorem{lemma}[theorem]{Lemma}

\newtheorem{remark}[theorem]{Remark}
\title{Searching  Lattice Data Structures of Varying Degrees of Sortedness
\thanks{Published in the International Journal of Data Structures, 1(1) 64-76 (2015)}}

\author{Mohammad Obiedat}
\affil{Department of Science, Technology, and Mathematics, Gallaudet University,
800 Florida Avenue NE,  Washington,  DC 20002-3695, USA\thanks{Email address: mohammad.obiedat@gallaudet.edu}}

\begin{document}
\maketitle

\begin{abstract}
Lattice data structures are space efficient and cache-suitable
data structures. The basic searching, insertion, and deletion
operations are of time complexity $O(\sqrt{N})$.  We give  a
jump searching algorithm of time complexity $O(J(L)\log(N))$, where $J(L)$ is the jump factor
of the lattice. $J(L)$  approaches $4$ when the degree of sortedness of the lattice approaches
$\sqrt{N}$.  A sorting procedure  of time complexity $O(\sqrt{N})$ that can be used, during the system idle time, to increase the degree of sortedness of the lattice is given.

\vspace{0.5cm}

\end{abstract}

\section{Introduction} \label{intro}

Searching is one of  the most time-consuming operation of many software applications.
Basically, a searching procedure takes a target key and it has to decide
whether that key is present or absent.  This seemingly simple
operation becomes  very challenging when the set of keys is
dynamically changing, by insertion and deletion, or when using minimum space and energy are very crucial.
Therefore, any data structure  that
is designed to organize pieces of information in a dynamic
environment  should be evaluated by  the space required to hold the keys, by the time required to perform the
searching, insertion, and deletion operations, and  by the suitability of the structure for caching or more generally for memory hierarchies.

In  Bjarne Stroustrup's recent paper \cite{Str12}, he   emphasizes the importance of   compactness and predictable memory access patterns for efficiency.  Stroustrup points out that power consumption  is roughly proportional to the number of memory accesses. Consequently,  the less data we store and move the better for efficiency, especially in infrastructure software and applications for hand-held devices like smartphones.  Stroustrup  makes a strong case in favor of using arrays instead of linked lists or more generally pointer-based structures. There are two important differences between  arrays and linked lists that  affect  the efficiency of their operations. First, arrays are more space efficient than linked lists. Second, arrays are more suitable for caching than linked lists, because an array's elements are stored in contiguous memory locations, while   a linked list's nodes are scattered in memory.

Most of the current index  structures are either pointer-based or  use extra storage space in order to obtain the
optimal  $O(\log{(N)})$ time complexity of their operations.
Obviously, such data structures are not suitable for
memory-constrained  systems. Sequential search of unsorted
arrays is space efficient and very suitable for caching, but unfortunately it is of time
complexity $O(N)$, which is  too slow for most applications.

The lattice data structure (LDS) was  novelly  introduced by Berkovich in 1992, see \cite{Ber92}. The space required to hold a lattice with $N$ keys is equal to the space required to hold an array with $N$ keys. Additionally,  it is possible to devise  more efficient algorithms using the LDS  than using the  array. This makes the LDS a good contribution to the current trends of developing data structures and algorithms for memory-constrained applications. Also, the LDS  operations have good  temporal and spatial locality which makes it  a rich environment for the ongoing research on cache-conscious data structures and cache-oblivious algorithms. Consequently, in addition to the theoretical importance of the LDS, we argue that this structure is the ideal choice for memory-constrained  systems, for real-time systems, for systems where power consumption is crucial, and for systems where incremental adjustment is feasible.

The basic searching, insertion, and deletion operations of the  LDS are
of time complexity $O(\sqrt{N})$. In this paper,  we give  a general
jump searching algorithm of time complexity $O(J(L)\log(N))$, where $J(L)$ is the jump factor
of the lattice. The jump factor of the lattice approaches $4$ when the degree of sortedness of the lattice approaches
$\sqrt{N}$.  In order to keep the jump factor of the lattice small,  we  provide a sorting procedure  of time complexity $O(\sqrt{N})$ that can be used, during the system idle time (e.g.,  by giving it the least priority when scheduling the lattice operations),  to increase the degree of sortedness of the lattice.

Experimental evaluation of randomly built LDSs shows  that the performance of the basic searching procedure of the LDS is similar to skip list search when the structures have up to $1000$ keys, and  the performance of the jump searching procedure  is similar to skip list search when the degree of sortedness of the lattice is close to $0.9h$, where $h$ is the height of the lattice.
When the ratio of the number of times the sorting procedure is called to the number of times the insertion and deletion operation are called is close to $5$, the performance of the jump searching procedure  is similar to its performance on
a lattice with degree of sortedness close to $0.9h$.

To obtain  a good performance of  the searching operation, a hybrid searching algorithm  can be implemented, together with the sorting procedure,   by searching the lattice using the  basic searching  procedure  when the degree of the sortedness of the lattice is less than or equal to $0.9h$, and the  jump searching procedure  when the degree of the sortedness of the lattice is greater than $0.9h$.

The layout of this paper is as follows.  In Section~2, we formalize
the definition of the LDS, present the basic searching, insertion, and
deletion algorithms,  and  then analyze the space and time complexities of
implementing these algorithms. In Section~3, we devise  a
general jump searching algorithm that can be used instead of the basic searching algorithm
to obtain a better performance and time complexity when the jump factor of the lattice is small. Then, we  provide a detailed analysis of the jump factor and devise a sorting algorithm that can be used to reduce the jump factor of the lattice during the system idle time.  In Section~4, we compare the performance of  the jump searching algorithm with  skip lists search. Then, we examine the effect of using the sorting algorithm on the jump factor of the lattice during the system idle time.
Conclusion  and recommendations for further research are given in Section~5.

\section{Lattice Data Structures} \label{sec:2}

In this section,  we formally define the lattice data structure and discuss
its basic algorithms and properties.

In order to define the lattice data structure, we first  recall the definition of Ferrers diagrams, see \cite{Yon07}.
Let $\lambda = (n_1\geqslant\cdots \geqslant n_m >0)$ be an integer partition of a positive integer $N$.
 A Ferrers diagram of shape $\lambda$ consists of $m$ rows, where the top row contains $n_1$  equisized square cells (hereafter referred to as cells), the second row from the top contains $n_2$ cells, etc. Each row is left-justified.  For a given non-negative integer number $h$, we are interested in up-side-down Ferrers diagrams of shape $(h+3,h+2,\ldots,2,1)$, see Figure~\ref{fig:LDSofHeightFour} for an up-side-down Ferrers diagram of shape $(7,6,5,4,3,2,1)$.  The following numbering schema and terminologies will be used for such  diagrams. Rows and columns  are numbered from bottom to top and from  left to right, starting with
one, respectively. Cells in each row and  each column are numbered
from left to right and from  bottom to top, starting with one,
respectively. By diagonal $k$, we mean the cells that lie on the
line segment that can be  formed by connecting the upper-left corner of the first cell  of row $k$  and
the lower-right  corner of the first cell  of column $k$ where $1
\leqslant k \leqslant h+3$. The head of diagonal  $k$ is the first cell of row $k$ and the tail is   the  first
cell of column $k$. Cells in each diagonal are numbered from head
to tail, starting with one. Using up-side-down Ferrers diagrams, instead of the usual Ferrers diagrams, is only a matter of convenience and preference.

\begin{definition} \label{defn:LDS}
A lattice data structure (LDS)  of height $h$ is an arrangement of a
finite set of distinct positive integers, referred to as  proper keys,  in an up-side-down Ferrers diagram of shape $(h+3,h+2,\ldots,2,1)$ such that:
\begin{enumerate}
\item   Each cell in row one contains  $0$.
\item   Each cell in column one contains  $0$.
\item Each cell in diagonal  $h+3$ contains $\infty$, except the head and the trail
      which contain $0$.
\item All other  cells contain proper keys, except the last $k$
      non-zero cells of  diagonal $h+2$ which  contain
      $\infty$ for some $0 \leqslant k \leqslant h-1$.
\item The proper keys in each row, column, and diagonal are in increasing order
      from left to right, from bottom to top, and from head to tail, respectively.
\end{enumerate}
\end{definition}

\begin{remark}
In the original definition of the LDS given by Berkovich~\cite{Ber92},
diagonals were not required to be sorted. As we will see in the next section, by requiring diagonals to be sorted one can apply a variety of  searching techniques to improve the time complexity  of the basic searching algorithm.
\end{remark}

Unless otherwise indicated, we will not distinguish between a cell's
location and its content, so we will say that ``cell two in row four
is less than cell three in row four'' instead of ``the content of
cell two in row four is less than the content of cell three in row
four''. A LDS of height $4$ and proper keys $\{3, 5, 6, 9, 12, 20, 30, 31\}$
is shown in Figure~\ref{fig:LDSofHeightFour}.

\begin{figure}[h]

\centering
\begin{picture}(170,145)(0,0)
\thicklines \put(40,0){\line(1,0){140}} \put(40,20){\line(1,0){140}}
\put(40,40){\line(1,0){120}} \put(40,60){\line(1,0){100}}
\put(40,80){\line(1,0){80}} \put(40,100){\line(1,0){60}}
\put(40,120){\line(1,0){40}} \put(40,140){\line(1,0){20}}

\put(40,0){\line(0,1){140}} \put(60,0){\line(0,1){140}}
\put(80,0){\line(0,1){120}} \put(100,0){\line(0,1){100}}
\put(120,0){\line(0,1){80}} \put(140,0){\line(0,1){60}}
\put(160,0){\line(0,1){40}} \put(180,0){\line(0,1){20}}

\put(47, 7){$0$} \put(67, 7){$0$} \put(87, 7){$0$}\put(107, 7){$0$}
\put(127, 7){$0$} \put(147, 7){$0$} \put(165, 7){$0$}

\put(47, 27){$0$} \put(47, 47){$0$} \put(47, 67){$0$}
\put(47,87){$0$} \put(47, 107){$0$} \put(47, 127){$0$}

\put(145, 27){$\infty$} \put(125, 47){$\infty$}
\put(105,67){$\infty$} \put(85, 87){$\infty$} \put(65,
107){$\infty$}

\put(67, 26){$3$} \put(87, 26){$6$} \put(105, 26){$31$} \put(125, 26){$\infty$}

\put(67, 47){$5$} \put(85, 47){$20$} \put(105, 47){$\infty$}
\put(67, 67){$9$} \put(85, 67){$30$}
\put(65, 87){$12$}

\end{picture}

\caption{A LDS of height 4}
\label{fig:LDSofHeightFour}
\end{figure}

\begin{remark}
The purpose of the improper keys in the boundary cells, row one, column one, and
diagonal $h+3$, is to serve as a wall; the LDS procedures can
not pass these cells and move down, to the left, or to the right.
Row one cells will serve as failure cells; if  a searching
procedure reaches these cells then it means that the key is absent.
Choosing the keys to be positive integers and the relation
$\leqslant $ to be the usual less   than or equal relationship is
only a matter of convenience; in practice the keys could  belong to
any set with a partial order relationship.
\end{remark}

In Figure~\ref{fig:LDSmovements}, we illustrate the eight possible
movements from a given cell $C$ to an adjacent cell. These
movements will be used in devising the  LDS  algorithms.

\begin{figure}[h]
\centering
\begin{picture}(200,125)
\put(60,60){\line(1,0){80}}\put(60,80){\line(1,0){80}}
\put(60,100){\line(1,0){80}}\put(60,120){\line(1,0){80}}

\put(60,60){\line(0,1){60}} \put(90,60){\line(0,1){60}}
\put(111,60){\line(0,1){60}} \put(140,60){\line(0,1){60}}

\put(65,65){DL} \put(97, 65){D} \put(115, 65){DR}

\put(73,85){L} \put(97, 85){C} \put(120, 85){R}
\put(65,105){UL} \put(97, 105){U} \put(115, 105){UR}

\put(30,40){U = Up, D = Down, L = Left, R = Right} \put(30,20){UL
= Up Left, UR = Up Right} \put(30,0){DL = Down Left, DR =  Down Right}

\end{picture}

\caption{Movements in a LDS from a given cell $C$}
\label{fig:LDSmovements}

\end{figure}

\begin{remark} \label{rem:LDSAdjacentCells}
If the cells adjacent to $C$ preserve the lattice structure, i.e.,
satisfy condition (5) of Definition~\ref{defn:LDS}, then $C$ also  preserves the lattice structure provided that $C> D$, $C> UL$, $C < U$, and $C<DR$ or $DR = 0$.
\end{remark}

Next, we describe the basic searching, insertion, and deletion algorithms of the LDS. We assume that we are given a LDS of height $h \geqslant 1$ and a key $K$ in the search space of the lattice.\\

\noindent \textbf{Searching.} The basic searching algorithm of the  LDS, SearchLDS, works as follows.
First, we set $C$ to be the  second  cell of diagonal $h+2$. Then we perform the following movements until $K$ is declared present or absent: If $K > C$, then  we move diagonally to $DR$ by setting  $C=DR$. If $K < C$, then we move down to $D$ by setting $C=D$. We declare $K$ is present if $C=K$, and declare $K$ is absent if $C = 0$. This basic searching algorithm, SearchLDS,  is given as Algorithm~{\ref{alg:SearchLDS}}.\\

\begin{algorithm}[h]

\caption{Basic Searching Algorithm of the LDS} \label{alg:SearchLDS}
\begin{algorithmic}
\STATE SearchLDS (LDS, $h$, $K$)

\STATE // Given a LDS of height $h$, and a key $K$.

\STATE // Determine whether $K$ is present or absent.

\STATE $C =$ second cell of diagonal $h + 2$;

\WHILE {($C\neq K$ \textbf{and} $C\neq 0$)}
     \IF {($K > C$)}
         \STATE $C = DR$;  // a $DR$ movement
     \ELSE
         \STATE $C = D$;  // a $D$ movement
     \ENDIF
\ENDWHILE
\IF  {$C = K$}
      \STATE $K$ is present at $C$;

\ELSE
       \STATE $K$ is absent;
\ENDIF

\end{algorithmic}
\end{algorithm}

\noindent \textbf{Insertion.} To insert a  key $K$ into a LDS, we first search the lattice for $K$.  If $K$ is present  then it cannot be inserted. If $K$ is absent,  then we find a temporary cell for $K$ as follows: If
diagonal $h+2$ has cells containing $\infty$, then we insert $K$ at
the first cell of diagonal $h+2$ that contains $\infty$. On the
other hand, if diagonal $h+2$ does not have any  cell that contains $\infty$,  then we expand the lattice by increasing its height by one and then insert $K$ in the second cell of diagonal $h+3$. After $K$ is inserted in the
temporary cell, say cell $C$, we maintain the lattice structure by
applying the inward algorithm, Inward,  given as  Algorithm~\ref{alg:Inward}.\\

\begin{algorithm}[h]
\caption{Inward Algorithm of the LDS} \label{alg:Inward}
\begin{algorithmic}

\STATE Inward (LDS, $C$)

\WHILE { ($C<D$ \textbf{or} $C<UL$)}

     \IF{ ($UL > D$)}
          \STATE swap $UL$ and $C$; // a $UL$ swap
     \ELSE
          \STATE swap $D$ and $C$; // a $D$ swap
     \ENDIF
\ENDWHILE
\end{algorithmic}
\end{algorithm}

\noindent \textbf{Deletion.} To delete a key $K$,  we first find its location,
say $K$ is located at  cell $C$. If $C$ is the last proper key of diagonal $h+2$, then we place $\infty$ in cell $C$. If $C$ is not the last proper key of diagonal $h+2$, then
we place the last proper key in diagonal $h+2$, say $F$, in cell $C$ and place $\infty$ in cell $F$. Finally, we maintain the lattice structure by applying the Inward algorithm if ($C<D$ or $C<UL$), and the  Outward algorithm, given as Algorithm~\ref{alg:Outward}, if ($C>U$ or  $C>DR$ and  $DR \neq 0$).
In all cases, we reduce the height of lattice by one if diagonal $h+2$ does not have any proper key.\\

\begin{algorithm}[h]

\caption{Outward Algorithm of  the LDS} \label{alg:Outward}

\begin{algorithmic}

\STATE Outward (LDS, $C$)

     \WHILE {($C>U$ \textbf{or} ($C>DR$ \textbf{and} $DR \neq 0$) )}

        \IF{ ($DR < U$ \textbf{and} $DR \neq 0$)}

            \STATE swap $DR$ and $C$;  // a $DR$ swap
        \ELSE
            \STATE swap $U$ and $C$;   // a $U$ swap
        \ENDIF
     \ENDWHILE
\end{algorithmic}
\end{algorithm}

In the following theorem, we give the number of proper and improper keys of a LDS of height $h$.

\begin{theorem} \label{thm:NumberProperImproperKeys}
Let $L$ be a lattice of height $h$, where  $h\geqslant 1$, and suppose that  diagonal $h+2$ contains $k$ proper keys where
$1\leqslant k \leqslant h$. Then
\begin{enumerate}
\item The number of proper keys in $L$ is $h(h-1)/2 + k$.
\item The number of improper keys in $L$ is $4h-k+6$.
\end{enumerate}
\end{theorem}

\noindent \textbf{Proof:} Straightforward.

\begin{corollary} \label{cor:LDSheight}
The smallest lattice that can hold $N$ proper keys is of height $h$ where
$h = \lfloor (1+\sqrt{8N-7})/2 \rfloor = \lceil (-1+\sqrt{8N+1})/2\rceil $.
\end{corollary}


In the following lemma, we give upper bounds on the number of comparisons to search  for a key, and the average number of comparisons to search  for a present key when a LDS is searched  using SearchLDS.

\begin{lemma} \label{lem:numberOfComparisonsInSearchLDS}
Let $L$ be a lattice of height $h$ and suppose that  diagonal $h+2$ contains $k$ proper keys where
$1\leqslant k \leqslant h$. If we search $L$ using SearchLDS, then
\begin{enumerate}
\item The number of comparisons to search for any absent key is $h+1$.
\item The number of comparisons to search for any present key is less than or equal to $h$.
\item The average number of comparisons to search for a present key is
 $$\frac{2h^3-2h+3k^2+3k}{3h^2-3h+6k}\approx \frac{2h}{3}.$$
\end{enumerate}
\end{lemma}
\noindent \textbf{Proof:} Straightforward.

\begin{theorem} \label{thm:timeCopmlexityBasicAlg}
The time complexity of  SearchLDS, Inward, and Outward algorithms
of  a LDS with $N$ proper keys is $O(\sqrt{N})$.
\end{theorem}
\textbf{Proof:} From  Lemma~\ref{lem:numberOfComparisonsInSearchLDS} and Corollary~\ref{cor:LDSheight}, the time complexity of SearchLDS is $O(\sqrt{N})$.
To find the time complexity of the Inward algorithm, suppose $C$ is located on diagonal $s$ and column $t$. Then with each $D$ swap, $s$ is reduced by $1$ and $t$ is  not changed, while with each $UL$ swap, $t$ is reduced by $1$ and  $s$ is  not changed. So with each swap,  $s+t$ is reduced by $1$ . Hence the Inward algorithm makes up to $2h$ swaps. Consequently,  it is an $O(\sqrt{N})$ algorithm. Similarly, the time complexity of the
Ouward algorithm is $O(\sqrt{N})$.

\begin{corollary} \label{cor:timeComplexityOfSearchInsertDelete}
If SearchLDS is used to locate a key or determine its absence, then the time complexity of the insertion and deletion procedures, described in this section,  of a LDS with $N$ proper keys is $O(\sqrt{N})$.
\end{corollary}

One of the main advantages of LDS over other  searching
structures  such as balanced trees (e.g., AVL trees \cite{ST12} and
self-adjusting trees \cite{ST85}), hash tables \cite{Knu73}, skip lists
\cite{Pug90_1,Pug90_2, Pug89},   Jumplists \cite{BCD03},  and J-lists
\cite{BLLSS89} is the space required to hold and maintain the
lattice. A lattice can be easily represented, either explicitly or implicitly,  by a one-dimensional array  where each cell is represented by one array element. In  explicit representation of LDSs, a lattice is mapped  into a one-dimensional array by mapping diagonal one to the first array element, diagonal two to the second and third array elements, etc. Implicit representation of LDSs is similar to explicit representation, except that only cells with proper keys are mapped into array elements. While implicit mapping is a little more space efficient that explicit mapping, explicit mapping is the natural choice, because of its simplicity and its superiority in term of time.

In Table~\ref{tab:spaceRequirement}, we compare the extra space
required to hold  a proper key by various searching data
structures.

\begin{table}[h]

\caption{Required extra space per proper key  by various searching
structures}\label{tab:spaceRequirement}
\centering

\begin{tabular}{ll}

\hline
 Structure & Extra space per proper key \\

\hline

LDSs-Explicit mapping& approaches $0$ as  $h\rightarrow\infty$ \\
LDSs-Implicit mapping& $0$  \\
J-lists & $1$ integer\\

Skip lists &  an average of $4/3$  pointers\\

Jumplists  &  $2$   pointers and $1$ integer\\

AVL trees  &  $2$  pointers and $2$ bits\\

Red-black trees  &  $2$  pointers and $1$ bit\\

Splay trees &   $2$  pointers\\
Hash tables  &  variable\\

\hline
\end{tabular}
\end{table}

\section{Jump Searching of Lattice Data Structures}
\label{sec:3}

Lattice data structures outperform other data structures
such as balanced trees, self-adjusting trees, and skip lists in
terms of space, time bounds of some operations, and suitability for memory hierarchies.
However,  the basic searching algorithm of the LDS, of time complexity $O(\sqrt{N})$,  cannot compete with the searching algorithms of these data structures in terms of time, especially when the size of the search space is large.
In this section, using the fact that diagonals  and columns of LDSs are sorted,  we devise a jump  searching algorithm
that can be used instead of SearchLDS to speed up the search. The basic idea of the jump searching algorithm is to  make wide jumps on diagonals and columns instead of the sequential movements to adjacent cells in the basic searching algorithm.
We also devise a sorting algorithm that can be used, during the system idle time, to incrementally increase the degree of sortedness of the lattice, and hence to improve the performance of
the jump  searching algorithm.

Jump searching algorithms over portions of a sorted array
are studied in detail in \cite{Shn78}. In
Algorithm~\ref{alg:JumpSearchLDS}, JumpSearchLDS, we given a general
jump searching algorithm of the LDS that can be used with a
variety of jump searching procedures of sorted arrays (e.g., binary and
interpolation) to obtain a better performance and  time complexity than SearchLDS.

\begin{algorithm} [h]

\caption{Jump Searching Algorithm of the LDS} \label{alg:JumpSearchLDS}

\begin{algorithmic}

\STATE JumpSearchLDS (LDS, $h$, $K$)

\STATE // Given a LDS  of height $h$, and a key $K$.

\STATE // Determine whether $K$ is present or absent.

\STATE $C =$ second cell of diagonal $h + 2$;

\WHILE {($C\neq K$ \textbf{and} $C\neq 0$)}

    \IF {($K < C$)}
    \STATE{// a downward jump}
    \STATE{Let $C$ be the first cell in the current column such that $K \geqslant
           C$;}

     \ELSE
        \STATE{// a diagonal jump}
        \STATE{Let  $C$ be the first cell in the current diagonal such
            that $K \leqslant C$, set $C=0$ if there is no such cell;}

     \ENDIF
\ENDWHILE
\IF  {$C = K$}
      \STATE $K$ is present at $C$;

\ELSE
       \STATE $K$ is absent;
\ENDIF
\end{algorithmic}
\end{algorithm}

Unless otherwise indicated, from now on, we will assume that $L$ is a LDS of height $h\geqslant 1$, and $K$ is a key in the search space of $L$. For brevity, let $d$ stand for a $DR$
movement. We define the search path
of $K$, with respect to $L$, to be the sequence of movements ($d$ or
$D$) until SearchLDS terminates.

\begin{definition} \label{defn:JumpFactor}
The jump factor of $K$, with respect to $L$, is  the number of
blocks of  identical movements in the search path of $K$.
\end{definition}

We  denote the jump factor of $K$ with respect to $L$ by $J(K;L)$, or $J(K)$, when $L$ is understood. For example, if the search path of $K$ is $ddddDDDDDdd$ then $J(K) = 3$.
We define the jump factor of $L$ by $$J(L)=\textrm{max}\{ J(K):K \textrm{ is a key in the search space of  } L \}.$$

Let $P$ be the procedure that performs the jump steps in
JumpSearchLDS. So, $P$ will take a sorted array, a column
(respectively, a diagonal) with $m$ elements and  for a given key
$K$ will return the first element $F$ such that $K\geqslant F$
(respectively, the first element $F$ such that
such that $K\leqslant F$ or set $C = 0$ if there is no such $F$).
In the following theorem, we give the
time complexity of JumpSearchLDS where the time complexity of $P$ is  $O(t_P(m))$.

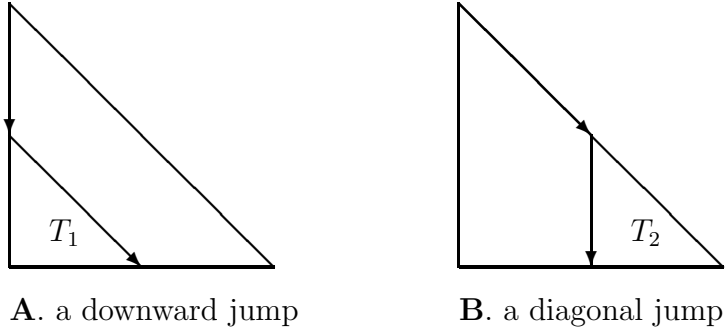
\begin{figure}[h]

\centering
\begin{picture}(300,140)

\put(10,10){\textbf{A}. a downward jump}

 \thicklines
 \put(10,30){\line(1, 0){100}}
 \put(10,30){\line(0, 1){100}}
 \put(10,130){\line(1, -1){100}}
 \put(10,130){\vector(0, -1){50}}
 \put(10,80){\vector(1, -1){50}}
 \put(25,40){$T_1$}

\put(180,10){\textbf{B}. a diagonal jump}
 \put(180,30){\line(1, 0){100}}
 \put(180,30){\line(0, 1){100}}
 \put(180,130){\line(1, -1){100}}
 \put(180,130){\vector(1, -1){50}}
 \put(230,80){\vector(0, -1){50}}
  \put(245,40){$T_2$}
\end{picture}

\caption{Jumps of the LDS}
\label{fig:LDSjumps}

\end{figure}

\begin{theorem}\label{thm:timeComplexityOfJumpSearchLDS}
 Let $L$ be a lattice of height $h$. If the time complexity of $P$ is $O(t_P(m))$,
 then  the time complexity of JumpSearchLDS is $O(J(L)t_P(h))$.
\end{theorem}
\textbf{Proof:} Given a lattice of height $h$. One can form a  triangle by connecting the following three points:
the lower-left corner of  the first cell of row one, the lower-right corner of the last cell of row one, and the upper-left corner of the last cell of column one. In Algorithm~{\ref{alg:JumpSearchLDS}}, when  a downward jump is
performed the keys in the  area outside triangle $T_1$ in
Figure~{\ref{fig:LDSjumps}-A} are eliminated from the search, and when a diagonal jump
is performed the keys in area outside triangle $T_2$ in
Figure~{\ref{fig:LDSjumps}-B} are eliminated from the search.
Since the time complexity of each jump is $O(t_P(h))$ and each search requires  up to $J(L)$ jumps, then the time complexity of JumpSearchLDS is $O(J(L)t_P(h))$.

\begin{corollary} \label{cor:timeComplexityOfBinaryJumpSearchLDS}
Let $L$ be a lattice with  $N$ proper keys. If the time complexity of $P$ is  $O(\log(m))$,  then  the time complexity of JumpSearchLDS is $O(J(L)\log(N))$.
\end{corollary}

The time complexity  of JumpSearchLDS depends on the jump factor of the lattice and on the procedure used to perform the jumps. From Theorems~{\ref{thm:timeCopmlexityBasicAlg} and \ref{thm:timeComplexityOfJumpSearchLDS}}, if  a binary-search-like procedure is used  to perform the jump steps, then the time complexity of
JumpSearchLDS is less than or equal  the time complexity of  SearchLDS provided that  the jump factor of the lattice is less than or equal to  $\sqrt{N}/\log(N)$. In the following lemma, we give  upper bounds on the jump factors of  keys.

\begin{lemma} \label{lem:upperBoundOfJ(K)}
Let $L$ be a lattice of height $h$, and $K$ be a key in the search space of $L$.
Suppose the last movement in the search path of $K$ ends on diagonal $s$ and column $t$.
Let $u = t - 2$ and $v = h - s + 2$, then

\begin{enumerate}
\item The number of $d$ movements in the search path of $K$ is $u$ and the number of $D$ movements is $v$.

\item

$
J(K)\leqslant \left\{
\begin{array}{ll}
u + v & \textrm{if $u = v$} \\
2u + 1  & \textrm{if $u < v$}\\
2v + 1 & \textrm{if $u > v$}
\end{array}
\right\}.
$
\item
$
J(K)\leqslant \left\{
\begin{array}{ll}
h-1 & \textrm{if $K$ is present} \\
h  & \textrm{if $K$ is absent}
\end{array}
\right\}.
$
\end{enumerate}
\end{lemma}
\textbf{Proof:}  Straightforward.

\begin{theorem}\label{thm:upperBoundsOnAverageJumpFactor}
Let $L$ be a lattice of height $h$. Then
\begin{enumerate}
 \item  The average of the jump factors of the present keys is less than or equal to $h/3 + 5/6 \approx h / 3$.
 \item The average of the jump factors of the absent keys is less than or equal to $h/2 + 1/2 \approx h / 2$, provided that  the probability  the search for a randomly chosen  absent key will terminate in a cell on row $1$ and column $t$  for  each $2\leqslant t \leqslant h+2$ is $1/(h+1)$.
\end{enumerate}
\end{theorem}
\textbf{Proof:} Follows from  Lemma~\ref{lem:upperBoundOfJ(K)} and  by using basic algebraic summations.

\begin{definition} \label{defn:DegreeOfSortedness}
Suppose $\alpha \in \{3, \ldots,h\}$. A LDS is said to be sorted of degree $\alpha$ (for brevity, $\alpha$-sorted) if the first proper key of diagonal $s$ is greater than the last proper key of diagonal $s-1$ for each
$4\leqslant s \leqslant \alpha + 2$.
\end{definition}
For example, the lattice in Figure~\ref{fig:LDSofHeightFour} is $3$-sorted but it is not $4$-sorted.

\begin{remark}
If $L$ is  $\alpha$-sorted,  then  the proper keys in the first $\alpha + 2$ diagonals are completely sorted. Also, if  $\beta = \alpha / h$ then the percentage of the proper keys in the first $\alpha + 2$ diagonals  is $\beta^2+(\beta-\beta^2)/(h+1)\approx \beta^2$. For example, if $\alpha = h/2$ then approximately $25\%$ of the proper keys in the lattice are completely sorted.
\end{remark}

\begin{lemma} \label{lem:upperBoundOfJ(K)alphaSorted}
Suppose $L$  is $\alpha$-sorted and $K$ is a key in the search space of $L$.

\begin{enumerate}
\item If $K$ is present and the last movement in the search path of  $K$ ends on diagonal $s$,
where $s<\alpha + 2$, then $J(K)\leqslant 2$.
\item If $K$ is absent and  the last movement in the search path of  $K$ ends on diagonal $s$,
where $s<\alpha + 1$, then $J(K)\leqslant 4$.
\end{enumerate}
\end{lemma}
\textbf{Proof:} (1) Suppose $K$ is present and $s<\alpha + 2$. Since the first proper key in diagonal $\alpha + 2$
is greater than the last proper key in diagonal $\alpha + 1$, then $K$ is less than the first proper key in diagonal
$\alpha + 2$. So, by a diagram chasing, the search path of $K$ is either
$D\cdots D$ or $D\cdots D d\cdots d$. Hence, $J(K)\leqslant 2$.\\
(2) Suppose $K$ is absent and $s<\alpha + 1$. By a diagram chasing, the search path of $K$ is either
$D\cdots D$, $D\cdots D d\cdots d$, $D\cdots D d\cdots dD$, or $D\cdots D d\cdots dD d\cdots d$. Hence, $J(K)\leqslant 4$.

\begin{theorem}\label{thm:upperBoundsOnJumpFactorSortedLDS}
If $L$  is $\alpha$-sorted, then $J(L)\leqslant\max\{4, 2h-2\alpha+2\}$.
\end{theorem}
\textbf{Proof:} Follows directly from Lemma~\ref{lem:upperBoundOfJ(K)alphaSorted}.

\begin{corollary} \label{cor:alphaSortedtimeComplexityJumpSearchLDS}
Let $L$ be an $\alpha$-sorted lattice of height $h$. If the time complexity of the procedure that performs the jump steps in JumpSearchLDS is $O(t_P(m))$, then the time complexity of JumpSearchLDS is $O(\max\{4, 2h-2\alpha+2\}t_P(h))$.
\end{corollary}

\begin{corollary}\label{cor:upperBoundsOnJumpFactorSortedLDS}
If $L$  is $h$-sorted, then
\begin{enumerate}
\item $J(K)\leqslant 2$ for any present key $K$, and
\item $J(K)\leqslant 4$ for any absent key $K$.
\end{enumerate}
\end{corollary}

\begin{corollary} \label{cor:hSortedtimeComplexityOfBinaryJumpSearchLDS}
Let $L$ be an $h$-sorted  lattice with  $N$ proper keys. If the time complexity of the procedure that performs the jump steps in JumpSearchLDS is $O(\log(m))$,  then  the time complexity of JumpSearchLDS is $O(\log(N))$.
\end{corollary}

In the following theorem, we give upper bounds on the average of the jump factors of the present and absent keys
for  $\alpha$-sorted lattices.

\begin{theorem}\label{thm:upperBoundsOnAverageJumpFactorAlphaSorted}
Suppose $L$  is $\alpha$-sorted and $\beta = \alpha / h$. If $0.5\leqslant \beta  \leqslant 1$, then
\begin{enumerate}
\item The average of the jump factors for the present keys is less than or equal to
$$\frac{(2\beta^3-4\beta^2+2\beta)h^2-(3\beta^2-6\beta+1)h+\beta - 3}{h+1}$$
$$\approx (2\beta^3-4\beta^2+2\beta)h-(3\beta^2-6\beta+1).$$
\item The average of the jump factors for the absent keys is less than or equal to
$$\frac{(\beta^2-2\beta+1)h^2+4h-4}{h+1}$$
$$ \approx (\beta^2-2\beta+1)h+4,$$
provided that  the probability  the search for a randomly chosen  absent key will terminate in a cell on row $1$ and column $t$  for  each $2\leqslant t \leqslant h+2$ is $1/(h+1)$.
\end{enumerate}
\end{theorem}
\textbf{Proof:}  To prove this theorem, we only need to use  Lemmas~{\ref{lem:upperBoundOfJ(K)}, \ref{lem:upperBoundOfJ(K)alphaSorted}},  and basic algebraic summations,  then substitute $\alpha  = \beta h$.\\

\begin{remark}
The functions $f(\beta) = 2\beta^3-4\beta^2+2\beta$ and
$g(\beta)=\beta^2-2\beta+1$ are decreasing on $[0.5, 1]$ with $f(0.5)=g(0.5)=0.25$ and $f(1)=g(1)=0$. So, from Theorem~\ref{thm:upperBoundsOnAverageJumpFactorAlphaSorted}, when $\beta$ approaches $1$, the average of the jump factors for the present keys approaches $2$ and the average of jump factors for the absent keys approaches $4$.
Also, from  Theorem~\ref{thm:upperBoundsOnJumpFactorSortedLDS}, $J(L)\leqslant\max\{4, (2-2\beta)h+2\}$.
\end{remark}

From Theorem~\ref{thm:upperBoundsOnAverageJumpFactorAlphaSorted}, the average of the jump factors of a lattice data structure  varies inversely with the degree of sortedness of the  lattice. So, in order for JumpSearchLDS to compete with searching structures of time complexity $O(\log(N))$,  we need to  maintain  LDS with high degrees of sortedness. In Algorithm~{\ref{alg:SortLDS}}, SortLDS,  we provide a procedure  for incrementally increasing the degrees of sortedness of LDS. SortLDS can be used during the system idle time to increase the degree of sortedness of the lattice.

\begin{algorithm} [h]
\caption{Sorting Algorithm of the LDS}
 \label{alg:SortLDS}
\begin{algorithmic}

\STATE SortLDS($T$, $h$)

\STATE // Given a LDS $T$ of height $h$.

\STATE // Swap the keys in the first two cells that prevent  $T$ from being $h$-sorted.

\STATE $i=4$;

\WHILE {($i\leqslant h + 1 $ \textbf{and}  the last proper key on diagonal $i$ is less than the
first proper key on diagonal $i+1$)}
  \STATE i = i + 1;
\ENDWHILE
\IF  {$i = h+2$}
      \STATE  break; // The lattice is $h$-sorted

\ELSE
      \STATE let $C$ be  the second cell of diagonal $i+1$
      \STATE let $F$ be the first cell of diagonal $i$ with key larger than the key in $C$;
       \STATE swap the keys in $F$ and $C$;
       \STATE Outward($T$, $C$);

\ENDIF
\end{algorithmic}
\end{algorithm}

In the following theorem we give the time complexity of SortLDS.

\begin{theorem}\label{thm:timeComplexityOfSortLDS}
The time complexity of  SortLDS algorithm
of  a LDS with $N$ proper keys is $O(\sqrt{N})$.
\end{theorem}
\textbf{Proof:} Straightforward.

\section{Experimental  Evaluation of LDSs and Skip Lists}\label{sec:4}

In this section, we  examine  the performance of the LDS searching algorithms for lattices with varying jump factors.
We compare the time performance values of SearchLDS,  JumpSearchLDS, and skip lists search. At the end of the section, we examine the effect of using SortLDS on the jump factor of the lattice during the system idle time.

A skip list is a probabilistic searching data structure, invented
by William Pugh in the late 1980's \cite{Pug90_1, Pug89}. Skip lists have
most of the desirable properties of balanced trees such as AVL trees
and self-adjusting trees with $O(\log(N))$ average time for most
operations,  yet they are more simple and  space efficient. The
rationale for comparing LDSs with skip lists is twofold. First,
skip lists are among the newest searching data structures, they are
compared against AVL trees and self-adjusting trees in
\cite{Pug89}, so by comparing LDSs with skip lists we gain insight
into the relationship between these important structures and LDSs.
Second,  the most important advantages of LDSs
are their simplicity and space efficiency, but these are claimed to
be the most important advantages of skip lists too.

The LDS algorithms are implemented in C programming language using Microsoft Visual Studio 2010 Compiler.
Skip lists source C code is obtained from William Pugh by anonymous ftp to ftp.cs.umd.edu.
The  proper keys are randomly chosen from the set [1, RAND\_MAX] = [1, 2147483647] and
stored  in  a file. Then the file is used to build and search lattices of varying heights and jump factors. Same keys are used to build and search skip lists. A binary-search-like procedure is used to perform the jump steps in JumpSearchLDS. The time  performance values  are the CPU time, measured in microseconds, on Intel(R) Core(TM) i5-2400 CPU @ 3.10GHz with 4.0GB RAM computer. Carrying out the experiments using the GNU C compiler on a Unix environment shows similar comparable results.

 From Lemma~\ref{lem:upperBoundOfJ(K)}, the number of each of the  $d$  and $D$ movements in the search path for any key depends only on the location where the search path terminates. So, the performance of SearchLDS is independent of the jump factor of the lattice. In Table~\ref{tab:SkipListsSearchLDSAndJumpSearchLDS}, we compare the performance of SearchLDS, skip lists search,  and JumpSearchLDS for lattices with varying jump factors. The time is the average searching time of the present keys, $h$ is height of the lattice, and $N$ is the number of keys. As we see from the table, the performance of JumpSearchLDS   is similar to SearchLDS  when $\beta = 0.8$, JumpSearchLDS  starts to perform better than SearchLDS  when $\beta \geqslant 0.9$, that is when more than 81\% of the keys are completely sorted. On the other hand, the performance of JumpSearchLDS   is similar to skip lists search   when $\beta = 0.9$, JumpSearchLDS  starts to perform better than skip lists search  when $\beta \geqslant 0.95$, that is when more than 90\% of the keys are completely sorted. When $\beta = 1$ the lattice is completely sorted and searching using JumpSearchLDS  is similar to searching a sorted array using binary search.

\begin{table}[h]
\caption{Timings  of SearchLDS, skip lists search, and JumpSearchLDS}

\centering

  \begin{tabular}{|l|l|l|l|l|l|l|l|}
\hline
$h$& $N$ & SearchLDS & Skip List & \multicolumn{4}{|c|}{JumpSearchLDS } \\
\cline{5-8}

& & & & $\beta = 0.8$ & $\beta = 0.9$ & $\beta = 0.95$ & $\beta = 1.0$\\
\hline

10& 55&0.048&0.059&0.064&0.061& 0.061 & 0.060\\
\hline
50& 1275&0.190&0.144&0.215&0.155& 0.134 & 0.128\\
\hline
100& 5050&0.347&0.207&0.370&0.223& 0.176 & 0.157\\
\hline
500& 125250&1.840&0.481&1.686&0.678& 0.357 & 0.228\\
\hline
1000& 500500&3.990&0.780&3.600&1.284& 0.557 & 0.267\\

\hline
\end{tabular}
\label{tab:SkipListsSearchLDSAndJumpSearchLDS}
 \end{table}

If the lattice is not $h$-sorted, then $\sqrt{N}$ calls to SortLDS increase the degree of sortedness of the lattice by at least 1, and hence reduce the average of the jump factors of the present keys of  the lattice. So,  using SortLDS during the system idle time will incrementally adjust the lattice, by increasing its  degree of sortedness, and hence improves the performance of JumpSearchLDS.

In order to simulate the performance of SortLDS during the system idle time,  we will assume the following scenario: the  system starts with an $h$-sorted lattice with $N$ proper keys,  when the system is not idle the  operations search, insert, and delete are performed,  when the system is idle SortLDS is repeatedly called, the system goes through idle time when $0.1N$ new keys are inserted, and the ratio of the insert operations to the delete operations is equal to $2$. The purpose of this hypothetical  scenario is to give us a rough idea about the impact of using SortLDS, during the system idle time, on the average of the jump factors of the present keys of the lattice. Let $\gamma$ to be the ratio of the number of times SortLDS is called to  the number of times the  insert and  delete operations  are called. Since the time complexity of the insert, delete, and SortLDS  procedures is $\sqrt{N}$, then, in Table~\ref{tab:AverageJumpFactorSortLDS}, instead of calculating the average of the jump factors of the present keys of the lattice with specific  ratios of  the idle time to the time of  insert and  delete operations, we will calculate the average of the jump factors of the present keys for the lattice with varying values of $\gamma$.
\begin{table}[h]

\caption{Average of the jump factors of the present keys for incrementally adjusted lattices using SortLDS}
\centering

\begin{tabular}{|l|l|l|l|l|l|}
\hline
$h$&  \multicolumn{5}{|c|}{Average jump factor with varying  $\gamma$ } \\
\cline{2-6}

&$\gamma = 0$ & $\gamma = 3$  & $\gamma = 4$  & $\gamma = 5$ & $\gamma = 6$\\
\hline

10& 2.2 & 1.82 & 1.77 & 1.77 & 1.77\\
\hline
50& 5.16 & 3.86 & 2.95 & 2.11 & 1.96\\
\hline
100& 8.03 & 5.81 & 4.14 & 2.75 & 1.97\\
\hline
500& 29.11 & 20.47 & 14.27 & 8.02 & 2.43\\
\hline
1000& 52.47 & 37.46 & 25.66 & 13.87 & 3.11\\
\hline
\end{tabular}
\label{tab:AverageJumpFactorSortLDS}
 \end{table}

 In Table~\ref{tab:AverageJumpFactorAlphaSorted},  we calculate  the average of the jump factors of lattices of varying heights and degrees of sortedness.

\begin{table}[h]

\caption{Average of the jump factors of the present keys for lattices with varying degrees of sortedness}
\centering

\begin{tabular}{|l|l|l|l|l|}
\hline
$h$&  \multicolumn{4}{|c|}{Average jump factor with varying  $\beta$ } \\
\cline{2-5}

&$\beta = 0.8$ & $\beta = 0.9$  & $\beta = 0.95$  & $\beta = 1.0$ \\
\hline

10& 1.93 & 1.75 & 1.64 & 1.64\\
\hline
50& 4.45& 2.70 & 2.05 & 1.92\\
\hline
100& 7.26 & 3.53 & 2.38 & 1.96\\
\hline
500& 29.53 & 10.22 & 4.26 & 1.99\\
\hline
1000& 57.37 & 18.57 & 6.54 & 2.00\\
\hline
\end{tabular}
\label{tab:AverageJumpFactorAlphaSorted}
\end{table}

 From Tables~\ref{tab:AverageJumpFactorSortLDS}, \ref{tab:AverageJumpFactorAlphaSorted}, the average of the jump factors of the present keys for  incrementally adjusted lattices using SortLDS with $\gamma = 0$, respectively $\gamma = 5$, is close to the average of the jump factors of the present keys for sorted lattices with $\beta = 0.8$, respectively $\beta = 0.90$. So, using the above scenario, when the system starts $\beta = 1.0$, and searching using JumpSearchLDS is similar to searching a sorted array using binary search. As the operations insert and delete are performed, $\beta$ deceases from $1.0$ to $0.8$  and JumpSearchLDS slows down and becomes similar to SearchLDS. Using SortLDS  with $\gamma = 5$ increases $\beta$ from $0.8$ to $0.90$,  and  searching using JumpSearchLDS becomes similar to skip lists search.

\section{Conclusions}\label{sec:5}

In this paper, we  have  formalized the definition of the lattice data structure and presented its basic operations.
We have shown that the  LDS  is as space efficient as the array and it is very suitable for caching. The worst case time complexity of the LDS basic operations is better than the corresponding worst case time complexity of most of  other data structures operations. In order to make this structure more attractive in term of the average case time complexity of the searching procedure, we  have devised a general jump searching algorithm  of time complexity $O(J(L)\log(N))$. We have also provided a sorting algorithm of time complexity $O(\sqrt{N})$ that can be used to reduce $J(L)$ during the system idle time. One possible direction for further research  is to explore using randomized searching procedures where the search starts at different locations  of the lattice instead of the second cell of diagonal $h+2$. Another possible direction is to perform a comprehensive experimental comparison of the LDS   with $B^+$-Trees and $T$-Trees to find out the best suitable structure for indexing in main memory. In \cite{Obi15P2}, we showed that the LDS is a very robust structure  for the  order-statistic operations.\\

\noindent  \textbf{Acknowledgements.} I would like
to thank Simon Berkovich for his helpful comments and suggestions. I
am also thankful to Regina Nuzzo who carefully read the manuscript
of this paper.

\end{document}